\documentclass[twocolumn,numberedappendix]{emulateapj}
\usepackage{graphicx,url,longtable,natbib}
\newcommand{\Nrich}{N_{\rm 19.5}}
\newcommand{\Ns}{N_{\rm s}}
\newcommand{\Mr}{M_{\rm r}}
\newcommand{\Mh}{M_{\rm h}}
\newcommand{\Msun}{h^{-1}M_\odot}
\newcommand{\Log}{{\rm log}}
\newcommand{\Ln}{{\rm ln}}

\shorttitle{the brightest group galaxy}
\shortauthors{Shen et al.}

\begin{document}

\title{The statistical nature of the  brightest group galaxies}
\author{Shiyin Shen$^{1,2}$, Xiaohu Yang$^{1,3}$, Houjun Mo$^4$,Frank van den
Bosch$^{5}$, Surhud More$^{6}$}

\affil{$^1$ Key Laboratory for Research in Galaxies and Cosmology, Shanghai
  Astronomical Observatory, Chinese Academy of Sciences, 80 Nandan Road,
  Shanghai, 200030, China \\
$^2$ Key Lab for Astrophysics, Shanghai 200234\\
$^3$Center for Astronomy and Astrophysics, Shanghai Jiao Tong University,
  Shanghai 200240, China\\
$^4$ Department of Astronomy, University of Massachusetts, Amherst MA
  01003-9305, USA\\
$^5$ Astronomy Department, Yale University , Box 208101,
  New Haven, CT 06520-8101, USA\\
$^6$ Kavli Institute for Cosmological Physics at the University of Chicago
  5640, S. Ellis Avenue, Chicago, IL 60637, USA } \email{ssy@shao.ac.cn}

\begin{abstract}
  We examine the statistical properties of the brightest group galaxies (BGGs) using a
  complete spectroscopic sample of groups/clusters of galaxies selected from the Data
  Release 7 of the Sloan Digital Sky Survey. We test whether BGGs and other bright
  members of groups are consistent with an ordered population among the total
  population of group galaxies. We find that the luminosity  distributions of BGGs do not
  follow the predictions from the order statistics (OS). The average luminosities of  BGGs are
  systematically  brighter than OS predictions. On the other hand, by properly taking into
  account the brightening effect of the BGGs, the luminosity distributions of the second
  brightest galaxies are in excellent agreement with the expectations of OS.  The
  brightening of BGGs relative to the OS expectation is consistent with a scenario that the
  BGGs on average have over-grown about 20 percent masses relative to the other member
  galaxies. The growth ($\Delta M$) is not stochastic but correlated with the magnitude
  gap ($G_{1,2}$) between the brightest and the second brightest galaxy. The growth
  ($\Delta M$) is larger for the groups having more prominent BGGs (larger $G_{1,2}$) and
  averagely contributes about 30 percent of the final $G_{1,2}$ of the groups of galaxies.
\end{abstract}


\section{INTRODUCTION}

The brightest group/cluster galaxies (hereafter BGGs) \footnote{Throughout
  this paper, we do not distinguish clusters from groups and simply use the
  term brightest group galaxy to refer the brightest member in a cluster or a
  group.}  are typically red, old ellipticals located near the centers of
their dark matter host haloes. Because of their extreme brightness and
uniformity in luminosity, they can be used to trace the large scale structure
and to study galaxy evolution in massive haloes \citep{Postman95, Rozo10,
  Ascaso11, Wen11}. On the other hand, BGGs are also known to differ from
ordinary ellipticals in their more extended surface brightness profiles and in their deviations
from scaling relations obeyed by other ellipticals \citep[e.g.,][]{Bernardi07, Linden07, Liu08}.

In the framework of hierarchical structure formation in a $\Lambda$CDM cosmology,
galaxies build up their stellar mass through mergers with other galaxies, and through in-situ
star formation, fed by cold flows and/or cooling flows that deliver gas to the potential well
centers of their host haloes. In the absence of any environmental effects, BGGs should simply
be statistical extremes: their extreme luminosities/stellar masses are merely a consequence
of them being {\it defined} as the brightest/most massive galaxies in their group. Put
differently, from a {\it physical} point of view, there is nothing special about a BGG. However,
it is well known that environment does play an important role.  Of particular importance is
the distinction between central galaxies, defined as the galaxy in a host halo with the
minimal specific potential energy, and satellite galaxies, which are galaxies that orbit around
a central galaxy \citep[see e.g.][]{Bosch08,Weinmann09,Pasquali10}. Typically central
galaxies are expected to grow in mass by cannibalizing their satellites
\citep{Dubinski98,Cooray05} and by being the repositories of cooling flows [although AGN
feedback may prevent such gas from being turned into stars; e.g. \citet{Rafferty08}],
whereas satellite galaxies are subjected to a number of processes that quench star formation
(i.e., ram-pressure stripping, strangulation) and strip mass (i.e. tidal stripping). If these
`environmental' effects have a significant impact on the luminosities and/or stellar masses of
the galaxies, one might expect BGGs, which typically are central galaxies \citep[though not
always, see e.g.][]{Skibba11} to evolve into a distinct population.

There are still controversies concerning the statistical nature of BGGs. Early analysis
\citep{Peebles68, Geller76, Geller83} found that the luminosity distribution of BGGs is
consistent with an statistical extreme value population.  More recent analyses, using
statistical tests such as the asymptotic form of the BGG luminosity function
\citep{Bhavsar85} and the luminosity gap between the first and second ranked group
members \citep{TR77,Loh06,Lin10,Tavasoli11}, reached the opposite conclusion: BGGs are
distinct compared to the extreme value statistics (however, see \citet{Paranjape12}.  There
have also been investigations that attempt to reconcile the two results by considering two
populations for the BGGs \citep{Bhavsar89, Bernstein01}. Most of these analysis focused on
rich clusters, where the number of member galaxies is large but the total number of systems
is limited. More recently, \citet{DC11} applied an order statistics to a large sample of
luminous red galaxies (LRG)\footnote{LRGs are typically the dominant galaxies in dark
matter haloes   and may therefore be similar to the BGGs considered here.}  selected from
the Sloan Digital Sky Survey \citep[SDSS,][]{York00} Data Release 7 \citep[DR7,][]{DR7} and
concluded that LRGs can be viewed as an extreme value sample when they are binned in
redshift rather than in terms of the richness of their host group.  Using LRGs selected from
the SDSS and the SDSS-III Baryon Oscillation Spectroscopic Survey \citep{Schlegel09},
\citet{Tal12} found that the large luminosity gap between the LRG and the most luminous
satellite can be reproduced by sparsely sampling a Schechter function obtained for galaxies
in random fields, suggesting that LRGs obey extreme value statistics\citep[see
also][]{More12,Paranjape12,Hearin13a}.

In the present paper, we study the statistical properties of BGGs using an order statistics
analysis similar to that carried out by \citet{DC11}. The order statistics (hereafter OS) studies
the expected distribution of the $k$-th order (largest value) of a given quantity (e.g.
luminosity) among a sample (group) with $N$ members (galaxies) which are randomly
drawn from an underlying probability density distribution. The extreme value statistic
(hereafter EVS) thus corresponds to $k=1$ of the OS. Our analysis is based on galaxies in
individual groups selected from a large spectroscopic survey, the SDSS DR7. This allows us
to divide our group sample into different richness bins, and to study the luminosity
distribution of the member galaxies separately for each of the richness bins (Section
\ref{Sec_Data}). We simulate  statistical samples by building mock groups using the
observed group member distribution function, and compare the properties of the member
galaxies in a given rank between real and  simulated samples to examine whether the
luminosity distributions of BGGs and other highly-ranked member galaxies are consistent
with the OS populations (Section \ref{Sec_OS}). Simple models are then presented to
understand how BGGs may over-grow their masses relative to other members (Section
\ref{Sec_model}). The specialties of BGGs are further discussed in Section \ref{Sec_TR} using
the Tremaine-Richestone test\citep{TR77} . Finally we summarize our results and make
discussions in Section 6.

\section{The Data}\label{Sec_Data}

In this paper, we use the SDSS galaxy group catalogs of \citet{Yang07}, constructed using
the adaptive halo-based group finder of \cite{Yang05}. The groups are selected from the
DR7 version of the New York University Value-Added Galaxy Catalogue
\citep[NYU-VAGC,][]{Blanton05}. From NYU-VAGC, we select all galaxies in the Main Galaxy
Sample with redshifts in the range $0.01 \leq z \leq 0.20$ and with a redshift completeness
$c > 0.7$. The resulting SDSS galaxy catalog contains a total of $639,359$ galaxies, with a
sky coverage of 7,748 square degrees.

Because of the fiber collision effect in the SDSS observation, two fibers on the same plate
cannot be closer than 55 arcsecs and so a small fraction of galaxies (about 7 percent)
eligible for spectroscopy do not have spectroscopic measurements. As a result, three group
catalogs are provided in \cite{Yang07}, samples I, II and III. Sample I only includes the
galaxies with measured redshifts from the SDSS, whereas sample II are further supplied with
a small fraction of galaxies with spectroscopic redshifts from other redshift catalogs. In
Sample III, those galaxies without spectroscopic measurements are {\it assigned} redshifts
according to their nearest neighbors. To avoid  incompleteness, we use sample III as our
group catalog.  As pointed out in \cite{Yang07}, the added redshifts in sample III may cause
the number of members of some of the galaxy groups to be overestimated. However, since
we are mainly concerned with the brightest galaxies, this fiber-collision effect is not
expected to have a significant impact on our results. Indeed, we have tested that using
sample II leads to negligible changes in all of our conclusions.

For each group in the catalog, the model magnitudes are used for both luminosity and
stellar mass measurements. A characteristic luminosity $L_{19.5}$ and a characteristic stellar
mass $M_{19.5}$ are defined, respectively, as the total luminosity and total stellar mass of all
group members with $\Mr<-19.5$.  The $r$ band absolute magnitude $\Mr$ is calculated
from the SDSS galaxy model magnitude and $K+E$ corrected to redshift $z=0.1$. The host
halo mass of each group is then estimated using the abundance matching of $L_{19.5}$ or
$M_{19.5}$ with the halo mass according to the halo mass function given by \citet{Tinker08}
for spherical over-density $\Delta=200$. The cosmological parameters used for the halo
mass function are consistent with the 7-year data release of the WMAP mission:
$\Omega_{\rm   m}=0.275$, $\Omega_{\rm \Lambda}=0.725$, $h=0.702$, and
$\sigma_8=0.816$ \citep{WMAP7}, where the reduced Hubble constant, $h$, is defined
through the Hubble constant as $H_0=100h~{\rm km~s^{-1}~Mpc^{-1}}$.

The halo masses for the galaxy groups described above are cosmology and model
dependent. To reduce such dependence, we use instead a richness parameter $\Nrich$,
defined as the number of member galaxies with $\Mr<-19.5$, as a halo-mass proxy. As we
will show, this richness parameter also plays as a key parameter in the OS study.  Since
group members are identified from the SDSS spectroscopic galaxy catalog, which is
complete to $r\sim 17.77$, $\Nrich$ can be obtained for all the groups at redshift $z<0.09$.
For our analysis, we  use only groups with $z<0.09$, which therefore is a volume complete
sample. We exclude the groups with all their members fainter than $-19.5$ mag. Such
groups have no halo mass been estimated in \citet{Yang07} and  have $\Nrich = 0$ in our
definition.  We also eliminate all groups with $f_{\rm edge}<0.9$ to reduce boundary
effects, where $f_{\rm edge}$ is a measure for the volume of the group that lies within the
SDSS survey edges(see Yang et al. 2007 for details). The total number of finally selected
groups is 113,436 and the number of member galaxies with $\Mr<-19.5$ is 159,503.

\begin{figure}
\centering
\includegraphics[width=85mm]{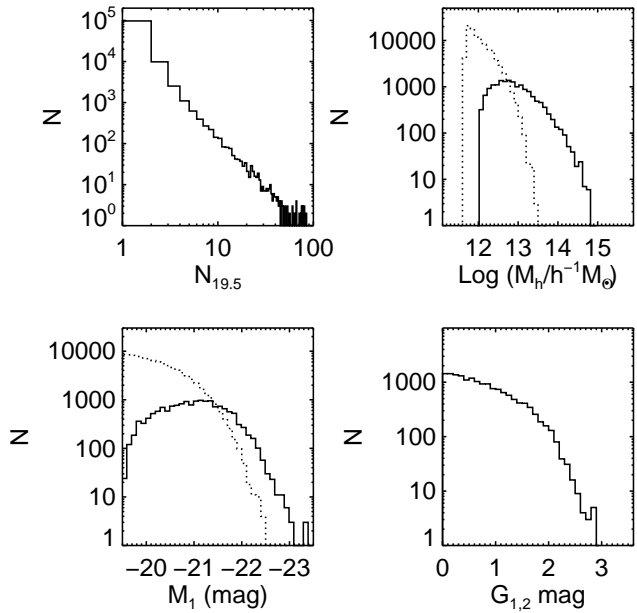}
\caption{The number histograms of the richness $\Nrich$, halo mass $\Mh$, BGG
  absolute magnitude $M_1$ and magnitude gap $G_{1,2}$ of our group sample.  In
the $\Mh$ and $M_1$ panels, the histograms of the single
galaxies($\Nrich=1$)  and  groups($\Nrich \ge 2$)  are shown by the dotted and
solid histograms separately. }
\label{Fig_Basic}
\end{figure}

We show the distributions of some of the basic parameters of our group sample in Fig.
\ref{Fig_Basic}. The top left panel   shows the histograms of group richness $\Nrich$.  Our
group sample includes a large population that only having one group member, i.e.
$\Nrich=1$. These single galaxies are the cases that no any other bright members can be
found inside their linking radii. In other words, all of their satellites were fainter than $-19.5$
mag. The number of such $\Nrich=1$ groups is 97,422. Besides them, our group sample
spans a very wide range, from poor groups ($\Nrich=2$) to rich clusters ($\Nrich > 100$).
The richest cluster in our sample has $\Nrich=144$ (Abell 2029). For $\Nrich \ge 2$ groups,
the host halo mass($\Mh$) ranges from $10^{12} \Msun$ to $\sim 10^{14.5}\Msun$.  Their
$\Mh$ distribution is shown as the solid histogram in the top right panel of \ref{Fig_Basic}.
In this panel, we also plot the $\Mh$ distribution of the single galaxies($\Nrich=1$) as a
separated dotted histogram. As we can see, the $\Mh$ of single galaxies also has a wide
distribution, about two orders of magnitude from $10^{11.6} \Msun$ to $\sim
10^{13.5}\Msun$. This wide $\Mh$ distribution of single galaxies stems from their wide
luminosity distribution, which is shown as  the dotted histogram in the bottom left panel of
Fig. \ref{Fig_Basic}.  In that panel, we also show the  BGG magnitude($M_1$) distribution of
the $\Nrich \ge 2$ groups as the solid histogram for comparison. The luminosity
distribution of the single galaxies has a monotonic shape, which can be approximated with a
Schechter function. On the other hand, the$M_1$  distribution of the $\Nrich \ge 2$ groups
shows a peaks at $\Mr\sim -21.7$ and has a very wide dispersion, spanning from our sample
limit $\Mr=-19.5$ to the very bright end $\sim -23.3$. As we will show in next sections, both
of the peak and dispersion of the BGG magnitudes are actually functions of $\Nrich$. The
bottom right panel shows the distribution of $G_{1,2}$, the magnitude gap between the
BGG and the second brightest group galaxy,

\begin{equation}
G_{1,2}=M_2-M_1 \,.
\end{equation}
This quantity plays an important role in the test of the statistical properties of the BGGs (see
Sections~\ref{Sec_gap} and Section \ref{Sec_TR}).

\subsection{The $\Mh$-$\Nrich$ correlation}

The halo mass ($\Mh$) we used is obtained by matching the number density of groups with
 the characteristic luminosity(the total luminosities of the $\Mr < -19.5$ group
members) above $L_{19.5}$   with the number density of dark matter haloes with masses
above $\Mh$ predicted by a halo mass function \citep{Yang07}. Not surprisingly, a strong
correlation between the group richness $\Nrich$ and $\Mh$ is expected.

\begin{figure}
\includegraphics[width=85mm]{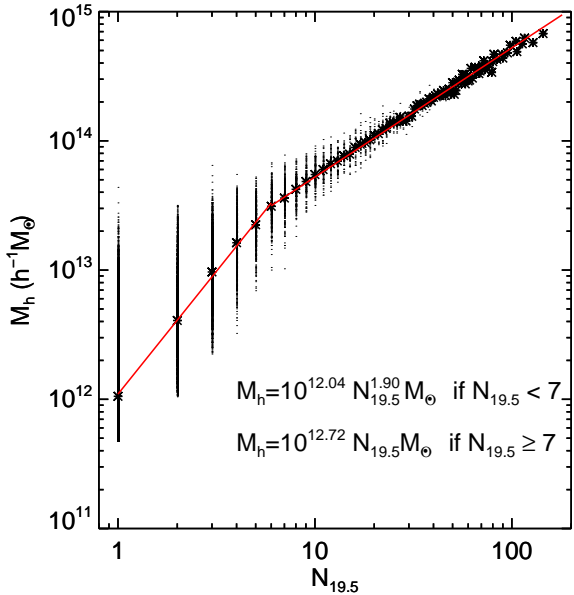}
\caption{The correlation between the group richness $\Nrich$ and the host halo
  mass $\Mh$.  The solid line shows the fitting relation of Equ.
  (\ref{Equ_NMh})} \label{Fig_NMh}
\end{figure}

We show the correlation between $\Nrich$ and $\Mh$ in Fig.  \ref{Fig_NMh}, where the
stars show the mean halo masses (in logarithm space) at each $\Nrich$.  In this plot, we see
again that $\Mh$  has a very large variance for single galaxies($\Nrich=1$).   With the
increasing of $\Nrich$,  the variance  of $\Mh$ at given $\Nrich$ decreases systematically.
This is because the variance of $L_{19.5}$(so that $\Mh$) decreases with the increasing of
$\Nrich$ for any given conditional luminosity function of the group members. The mean
$\log \Mh$ shows a linear correlation with $\log \Nrich$ for rich groups with $\Nrich \geq
7$. Such a linear relation has also been obtained in other studies where the halo masses are
estimated from independent measurements, e.g., satellite kinematics \citep{Becker07} and
the X-ray data \citep{Andreon10}. For poor groups with $\Nrich < 7$, however, the slope of
the $\Mh$ - $\Nrich$ relation becomes steeper, which is attributed to the different shapes
between the halo mass function and group member luminosity function at the low mass
end. For simplicity, we fit the $\Mh$-$\Nrich$ correlation with a broken-power law which
breaks at $\Nrich=7$. Fixing the power law index to be $n=1$ at $\Nrich \geq 7$ we find
that
\begin{equation}\label{Equ_NMh}
  \Mh=\left\{
\begin{array}{ll}
   10^{12.04}\, \Nrich^{1.90}\, \Msun &     \mbox{if $\Nrich < 7$}\\
   10^{12.72}\, \Nrich\,  \Msun       &     \mbox{if $\Nrich \geq 7$}
\end{array}
\right. \,.
\end{equation}
This is shown in Fig.  \ref{Fig_NMh} as the solid line. The scatter in the relation is quite small
for rich systems (about 0.2 dex at $\Nrich =7$), so that $\Nrich$ is a good indicator of halo
mass. For poorer systems, the scatter becomes increasingly larger. Because it has been
assumed that there is no scatter between $L_{19.5}$ and $\Mh$\citep{Yang07}, the scatter
of $\Mh$ at given $\Nrich$ actually reflects the scatter between $L_{19.5}$ and $\Nrich$.

Although $\Nrich$ and $\Mh$ have a  correlation as shown in  Equ. \ref{Equ_NMh}, they are
two different measurements of the global properties of groups. From statistical point of
view, our sample is complete in $\Nrich \ge 1$ while it is not in $\Mh \ge M_{\rm h,min}$,
where $M_{\rm h,min}$ is  the minimum $\Mh$ of our group sample, $ \sim 5\times
10^{11}\Msun$.  The reason is that there is no $\Mh$ estimation for these groups with all
their members fainter than $-19.5$ mag(i.e. $\Nrich=0$). In our group sample, $M_{\rm
h,min}$ corresponds to the case of a single galaxy with the minimal sample luminosity
$\Mr=-19.5$ mag. It is certain that there are $\Nrich=0$ groups with $\Mh \ge M_{\rm
h,min}$. Therefore, in following, we will use $\Nrich$ as the reference parameter of the
groups to make OS studies. When required, we will use Equ. \ref{Equ_NMh} to make
discussions on $\Mh$.

\subsection{Group member luminosity function}

To quantify the statistical nature of BGGs, the luminosity distribution of the group members
need to be determined. The group member luminosity function is shown to depend on the
host halo mass \citep{Hansen05, Weinmann06,   Yang08, Hansen09}.  Here, we present the
luminosity distributions of the group members in different richness bins.

For  $\Nrich=1$ groups,  the  group member luminosity distribution equals to the BGG
luminosity distribution and  we can not obtain any further conclusions from OS. Therefore,
we will limit our following studies only on $\Nrich \ge 2$ groups. For clarity, we will note
$\Nrich=1$ groups as `single galaxies' hereafter. When we state `group', we only refer to the
groups with  $\Nrich \ge 2$.

We separate  our groups into 6 $\Nrich$ bins: $[2, 2]$, $[3, 4]$, $[5, 6]$, $[7, 10]$, $[11,20]$,
$[21,144]$. The widths of the bins are a compromise between the resolution in $\Nrich$ and
the number of groups and group members within each bin. For reference, the average
logarithmic halo mass, the number of groups and group members are listed in Table 1 for
each of the richness bins.

\begin{figure}
\includegraphics[width=85mm]{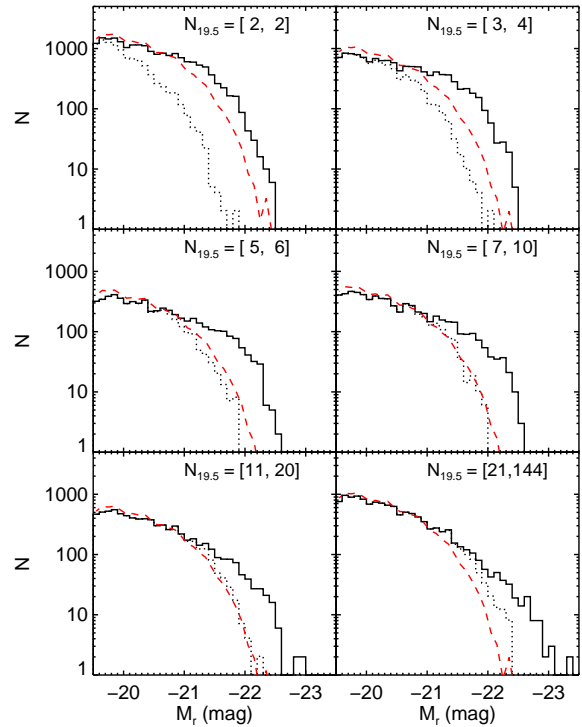}
\caption{The $r$ band absolute magnitude distributions of the group
members(solid histograms) and the satellites(dotted histograms) in  different $\Nrich$
bins. In each panel, the dashed line shows the luminosity
  distribution function of the single galaxies, which are   normalized to the
same number of
  the group members in each $\Nrich$ bin.}
\label{Fig_CLF}
\end{figure}

The conditional luminosity functions (hereafter CLF) of the group members in different
richness bins are shown as the solid histograms in Fig. \ref{Fig_CLF}.  To see whether the CLF
changes with group richness(halo mass),  we also plot in each panel the luminosity function
of the single galaxies as a reference(dashed line). In a given panel, this function  has been
normalized by the same number of the group members in each $\Nrich$ bin. As we can see,
the CLFs of the group members in all $\Nrich$ bins are  brighter than the luminosity
function of the single galaxies  and vary significantly  with $\Nrich$. Groups with increasing
$\Nrich$ contain systematically larger fraction of bright galaxies with $\Mr< -22$. The
Kolmogorov-Smirnov (KS) test probabilities that the CLFs follow the same distribution  are
close to zero for any two $\Nrich$ bins.

In each panel of Fig. \ref{Fig_CLF}, we also show the CLF of the satellite members, i.e.  the
$\Mr$ distribution of the group members except  BGGs, as the dotted histogram. Similar as
the CLFs of all group members, the satellite distribution functions also show a systematical
dependence on $\Nrich$. Satellites are also on average brighter in higher mass haloes.

In OS, the statistical properties of the $k$th order ($k$th brightest) members depend both
on the sample size $N$ ($\Nrich$ in our case) and the underlying distribution function of
the sample members \citep{DC11,Paranjape12}. In Fig. \ref{Fig_CLF}, we have shown that the
CLFs of the group members and the CLFs of the satellites both vary systematically with the
group richness (halo mass).  This demonstrates clearly that galaxies in these groups do not
have the underlying luminosity distribution as the general population. This also means that
BGGs cannot be tested as the extremes of a general galaxy population. \citet{Paranjape12}
showed that the group members follow a universal luminosity distribution. However, their
conclusions are only based on the comparison between two subsamples of groups with $N
\ge 10$ and $N \ge 15$ members.

Thus, in our following study of the statistical properties of the group members using OS, we
will consider groups in given $\Nrich$ bins, and use the CLF corresponding to the $\Nrich$
as the underlying distribution to test if BGGs are consistent with the extreme value statistics
of the galaxy population contained in such groups.

\section{The order statistic prediction}\label{Sec_OS}

In this section, we start from the observed CLFs  and make predictions for the statistical
properties of the BGGs and other bright members using OS. By comparing  the statistical
properties of the ordered members of the real groups with the OS predictions, we test
whether BGGs and other bright members are consistent with the OS.

\subsection{The brightest group galaxies}
\label{Sec_BGGOS}

We make OS predictions for  BGGs by building a sample of mock groups for each $\Nrich$
bin, which is designed to  have the same richness distribution and underlying luminosity
distribution as the real groups but with their  members assigned in a statistical (random)
way. To achieve this, we make mock groups that have exactly the same richness as the real
groups in the sample. We then assign random luminosities to the members of each mock
according to the observed  CLF (solid histogram in Fig. \ref{Fig_CLF}). By this construction,
the BGGs of the mock groups can be compared with the BGGs of the real groups. For clarity,
we refer to the mock groups as OS groups, and the BGGs of the OS groups as OS BGGs.

\begin{figure}
\includegraphics[width=90mm]{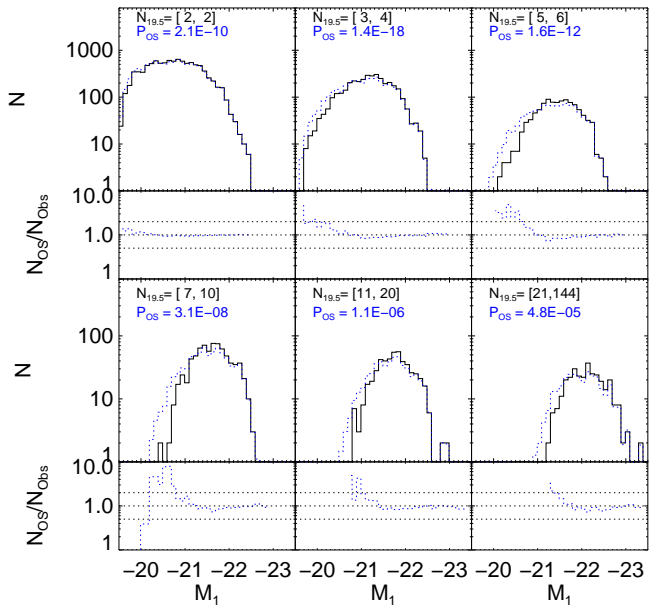}
\caption{The distributions of the $r$ band absolute magnitudes of the BGGs in
  six $\Nrich$ bins (as indicated in each of the six panels).
In the upper part of each panel, the solid
histogram shows the $M_1$ distribution  of the real  BGGs, while the dotted
histogram shows the $M_1$ distribution of the OS BGGs.  The KS test
probability ($P_{\rm OS}$) that the  OS BGGs follow the same $M_1$ distribution
as the  real
BGGs is labeled on top of each panel. In  the lower half panel,  we show
the ratio in the number between OS BGGs and real BGGs.  The
three dotted lines show the constant ratios, 0.5, 1 and 2 respectively. }
\label{OS_BGG}
\end{figure}

We show the  histograms of the absolute magnitude   of the real and OS BGGs  in Fig.
\ref{OS_BGG}. In each panel ($\Nrich$ bin), the solid histogram shows the real BGGs, while
the  dotted histogram represent the OS BGGs. We make a comparison of the $M_1$
distributions of the real BGGs and OS BGGs  with a non-parametric KS test. The KS test
probability that the real BGGs follow the same distribution as the OS BGGs ($P_{\rm OS}$) is
labeled on top of each panel. As we can see, for all richness bins, the KS test probabilities
that the OS BGGs follow the same distributions as the real BGGs are very small ($<10^{-4}$).

To further quantify how the $M_1$ distribution of the real BGGs deviates from the OS
prediction, we calculate the ratio of the two histograms and plot it in the lower part of each
panel. As one can see, the OS BGGs are systematically biased to  the low luminosity ends for
all $\Nrich$ bins. In other words, the real BGGs are systematically brighter than the OS
predictions. \emph{These systematical deviations are visually comparable or even larger in
high $\Nrich$ bins(lower panels) than in  low  $\Nrich$ bins(upper ones). However, because
of the numbers of the groups are less in large $\Nrich$ bins, the statistical significance there
is even less(larger KS test probabilities) .}

The  brighter luminosity of the real BGGs than the OS predictions may be consistent with the
evolutionary scenario of  BGGs.  For instance, the  `environmental' effects may boost the
luminosities of BGGs  through galaxy cannibalism, star formation in cooling flow etc. We will
come back to this scenario with simple models in Section \ref{Sec_model}.

\subsection{The second brightest group galaxies}\label{Sec_OS_SGG}

In this section, we further check whether the magnitude ($M_2$) distribution of the second
brightest group galaxies (hereafter SGGs) are consistent with the OS predictions. In the
above section, we have shown that the observed BGGs are not consistent with an OS
population, namely the luminosities of the BGGs must have evolved from the expected
statistical extremes. Because of this, we can not use the observed CLFs of the group member
as the underlying distribution to make OS predictions for the SGGs before  the change in
the BGGs is taken into account (see Section \ref{Sec_model}).

As an alternative, we may view  BGGs as a distinguished population(centrals) and exclude
them from other members(satellites). In this case, the SGGs can be named in another way,
the BSGs, the brightest satellite galaxies. Our statistical test is then changed to whether the
observed $M_2$ distributions are consistent with  the BSGs from OS prediction.

\begin{figure}
\includegraphics[width=85mm]{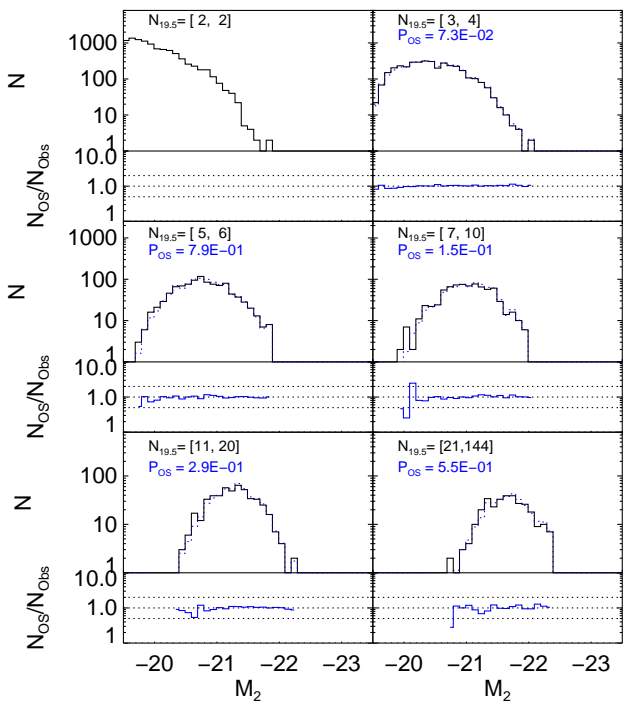}
\caption{The distributions of the $r$ band absolute magnitudes of the SGGs,
 i.e. the BSGs  in six  $\Nrich$ bins.  The solid histogram  shows the $M_2$
distribution  of the
 real  SGGs, while the dotted histogram shows that for the BSGs
obtained from the OS prediction.
The arrangement of the figure is the same as
Fig. \ref{OS_BGG}.} \label{OS_SGG}
\end{figure}

As  for BGGs,  we also generate a mock sample of satellite galaxies for each richness bin,
which have the same group richness distribution and follow the same satellite CLF as the
real groups (dotted histograms in Fig. \ref{Fig_CLF} ).  We then make OS predictions for the
luminosity distributions of the BSGs from this mock sample and compare them with the
$M_2$ distribution of real groups. The results are shown in Fig. \ref{OS_SGG}. The
arrangement of this figure is the same as Fig. \ref{OS_BGG}.

For the first richness bin, $\Nrich = 2$, there is only one satellite for each group, and so by
definition the mock BSGs will follow  the same luminosity distribution as the real SGGs. We
therefore do not show the distribution of the OS BSGs for this case. For all other richness
bins, the luminosity distributions of the real SGGs are in excellent agreement with the BSGs
obtained from the OS prediction. The KS test probabilities that the OS BSGs follow the same
distribution as the real SGGs are significant for all the richness bins. This result implies that,
unlike the central galaxies, the luminosities of the bright satellite galaxies have not been
significantly changed by `environmental' effects relative to the underlying population.

\section{Modeling  the magnitude distribution of BGGs}
\label{Sec_model}

As shown above, the luminosity distribution of  the BGGs are not consistent with the OS
population expected from the CLF, while the SGGs are consistent with being the extremes of
the satellite members expected from the CLF of satellite galaxies. These results suggest that
BGGs are distinct from other member galaxies. In the current hierarchical structure
formation model,   BGGs are typically associated with the host haloes and located at the
centers of their hosts, while other member galaxies are typically associated with dark matter
subhaloes. Thus, being the centrals,  BGGs may have the privilege to grow more during their
evolution through galaxy cannibalism or through star formation fueled by cooling flows.
Here, we test such a scenario using a toy model. We assume that the assembly history of a
halo  is only correlated with its halo mass, so that for groups with a given richness (halo
mass), their members follow the same initial luminosity distribution function. After the BGGs
settle into the group centers, they are assumed to be brightened by some environmental
processes, while the luminosities of other members are assumed to remain unchanged. We
denote the initial absolute magnitude of a BGG by $M^*_1$ and the subsequent change as
$\Delta M$ ($>0$), so that the final magnitude of the BGG $M_1$ is written as
\begin{equation}\label{Equ_M1}
M_1=M^*_1-\Delta M .
\end{equation}
The magnitudes of the other member galaxies remain unchanged, so that
\begin{equation}\label{Equ_M2}
M_k=M^*_k\,,
\end{equation}
where $k$ is the rank order in luminosity of group members.   In this scenario, the
distributions of $M^*_1$ and $M_k (k=2,\cdot\cdot\cdot\Nrich)$ are expected to follow the
OS predictions of the same underlying (initial) distribution function, if the model for $\Delta
M$ is correct. Thus, to test the model,   we apply the {\it reversed} brightening process
($\Delta M$) to the observed BGGs ($M_1$) and check whether or not the `dimmed' BGGs
($M^*_1$) is consistent with the OS prediction of the luminosity distribution of galaxies in
the sample where BGGs are replaced by their `dimmed' counterparts (referred to as the
`BGG-dimmed' sample). Moreover, to get further constraints on $\Delta M$, we may also
check whether the magnitude of SGGs ($M_2$) and galaxies in other rank orders are
consistent with the OS predictions based on the underlying distributions of the
`BGG-dimmed' sample.

In practice,  for each $\Nrich$ bin we reduce the luminosity of each BGG by an amount of
$\Delta M$, where $\Delta M$ is parameterized by some simple model parameters (to be
specified later). In order to ensure that the BGG retains its brightest status after dimming,
$\Delta M$ is constrained to be smaller than the observed magnitude gap, $G_{1,2}$, for
any given group (to be discussed later). After this dimming, we make statistical tests on
whether all the ordered members become being consistent with OS predictions. To do that,
we apply a shuffling algorithm on the group members of this `BGG-dimmed sample' within
each $\Nrich$ bin. More specifically, we first assign all member galaxies of the
`BGG-dimmed' sample in a given richness bin into an array, $\{i=1,2,\cdot\cdot\cdot,N_{\rm
gal}\}$, so that each group is represented by a set of integers equal to the indices assigned
to their member galaxies. We then randomly shuffle the elements of the corresponding
magnitude array $M_i$ so that each index is now associated with a new magnitude in the
shuffled array. Finally we build a sample of new groups so that each group contains member
galaxies specified by their original indices but with their magnitudes according to the
shuffled magnitude array. Because the members of each new group are assigned in a
statistical (random) way, the BGGs and  the members at any other ranks  can be considered
as a statistical population of the primordial distribution (i.e. the distribution before the
evolution of the BGGs).

For convenience, we refer these new groups as `shuffled groups'. Finally, we apply a
non-parametric KS test to check the probabilities that the distributions of the`BGG-dimmed'
sample, (e.g. $M^*_1, M_2$) are consistent with those of the shuffled groups.

\subsection{Average brightening of the brightest group galaxies}
\label{Sec_Cmodel}

As a first step, we estimate the average brightening of the BGGs relative to that of other
member galaxies. For a given $\Nrich$ bin, we assume  that the BGG brightening is
independent of any other group properties and can be parameterized by a simple
parameter $\Delta M$.  In principle, one can assume that the BGG brightening $\Delta M$
has some random scatter.However, we have found that our statistical model can not provide
any constraints on this scatter.  The reason is that any random scatters on the BGGs do not
change the statistical nature of the BGGs and are degenerated with the random shuffling
process. Therefore, the parameter $\Delta M$ we assumed in different $\Nrich$ bins is
better to be understood as the average values of the BGG brightening rather than a
constant. Nevertheless, we refer to this  BGG brightening model parameterized by a
constant $\Delta M$ as $C$ model below.

For a given $\Nrich$ bin and a given value of $\Delta M$, we calculate the  KS test
probabilities, $P_{\rm BGG}$ and $ P_{\rm SGG}$, which are  the $M^*_1$ and $M_2$
distributions of the BGG-dimmed sample follow the same distributions as those of the
shuffled groups respectively.  We  define a combined KS test probability,
\begin{equation}\label{Equ_BS}
 P_{\rm B,S}=P_{\rm BGG}*P_{\rm SGG}\, .
\end{equation}
For each $\Nrich$ bin, we then search the best values of $\Delta M$  in the parameter space
by maximizing $P_{\rm  B,S}$. To reduce the statistical fluctuations during the shuffling
process, we make the shuffling process 100 times for each assumed value of $\Delta M$,
and use the mean of these 100 realizations to estimate the KS test probabilities.

\begin{figure}
\includegraphics[width=85mm]{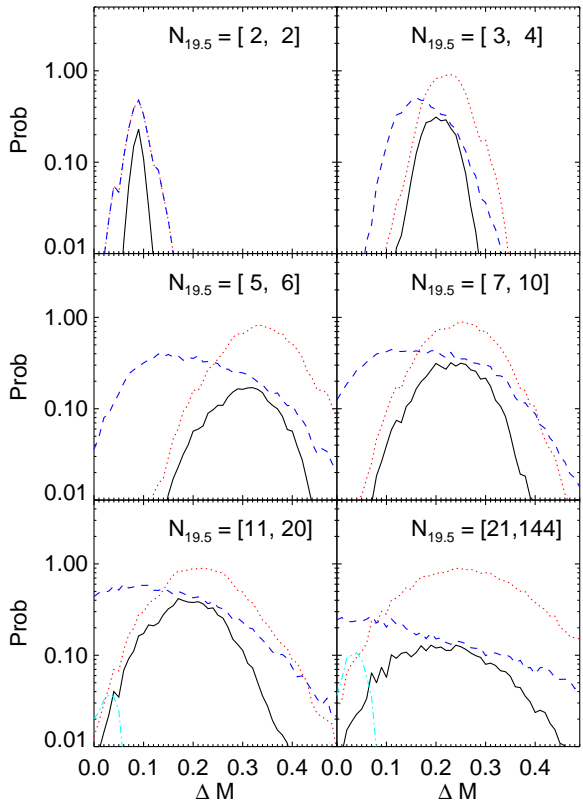}
\caption{The KS test probabilities $P_{\rm BGG}$, $P_{\rm SGG}$ and $P_{\rm B,S}$ as function
 of the average BGG brightening $\Delta M$  for 6 different $\Nrich$ bins, which are shown by
 the dotted, dashed and solid lines in each panel respectively.
 The dot-dashed lines in bottom two panels show the  KS test probability $P_{\rm GAP}$(see Section \ref{Sec_gap}).
 For the other $\Nrich$ bins, $P_{\rm GAP} < 0.01$  at all  $\Delta M$   values and therefore are not shown. }
\label{Fig_modelBS}
\end{figure}

In Fig. \ref{Fig_modelBS}, we show the $P_{\rm BGG}$, $P_{\rm SGG}$ and $P_{\rm B,S}$ as
function of $\Delta {M}$ for 6 different $\Nrich$ bins with the dotted, dashed and solid lines
respectively.  As one can see,  we get very good  estimates of $\Delta M$ (with $P_{B,S}  >
0.1 $)  in all richness bins. Although  the constraint on $\Delta M$ mainly comes from the
BGG distribution($P_{\rm BGG}$),  the behavior of the SGG distribution($P_{\rm SGG}$) also
shows good consistence.

For each $\Nrich$ bin, we define the best estimate of the average BGG brightening,
$\Delta\bar{M}$, as the value at which the accumulated probability starting from $\Delta
M=0$ is half of that from $\Delta M=0$ to $\Delta M=0.5$. This estimate matches the
maximum $P_{\rm BS}$ estimation well in most of the cases, but is more robust. The
uncertainties of the best estimates are then calculated from the 68 percentiles of the
likelihood distribution on each side of $\Delta\bar{M}$. The values of $\Delta\bar{M}$ so
obtained are $0.09$, $0.20$, $0.30$, $0.23$, $0.19$, $0.22$ magnitudes for  six richness
bins, respectively. They are plotted together with their uncertainties in the top panel of Fig.
\ref{Fig_DeltaM} and are listed in Table 1.

The actual average magnitudes we have used to dim the BGGs are slightly smaller than the
values of $\Delta\bar{M}$, because we have forced  $\Delta M < G_{1,2}$ so that each BGGs
remains to be the brightest after the dimming. However, since the fraction of groups with
$G_{1,2} < \Delta\bar{M}$   is quite small (see the bottom right panel of Fig. 1), the
difference between the real average dimming and $\Delta\bar{M}$ is small, typically $\sim
0.01$ mag.

We may abandon the constraint $\Delta M < G_{1,2}$  in our model. However, we find that
the value of $\Delta\bar{M}$ can no longer be strongly constrained by the observational
data.  The reason is, because then a galaxy at any rank can in principle become the BGG
when $\Delta M$ is chosen sufficiently large. For example, consider a case in which the
initial luminosities of the current BGGs are fainter than the third brightest group galaxies
(TGGs) of the current groups.  Thus, after the BGG-dimming, the SGGs of the current groups
become the BGGs of the initial groups and the TGGs of the current groups become the SGGs
of the initial groups. Since the distributions of the SGGs and TGGs of the current groups are
consistent with the OS prediction(Section \ref{Sec_OS_SGG}),  such a scenario is also
acceptable if we give up the assumption $\Delta M < G_{1,2}$.

Except the lowest $\Nrich=2$ bin,  the average BGG brightening  $\Delta\bar{M}$ is roughly
a constant $\sim 0.2$ mag. Using the $\Nrich-\Mh$ relation in Equ. \ref{Equ_NMh}, we label
the median $\Mh$ of the groups in  6 $\Nrich$ bins on the top axis of Fig. \ref{Fig_DeltaM}.
We see that the $\Nrich > 2$  groups   corresponds to the haloes with $\Mh > 10^{13}
\Msun$.

It is worthy to mention that the low $\Nrich$ groups are not complete in halo mass(see Fig.
\ref{Fig_NMh}). There are  low mass groups with all their members/satellites $\Mr$ fainter
than $-19.5$ mag. Such groups  have $\Nrich=0/1$  by definition. The BGGs of these
$\Nrich < 2$ groups may also over-grow their stellar masses through, e.g. cannibalizing the
very faint satellites($\Mr < -19.5$ mag). However, our OS studies requires at least two
members for each group and can not be applied to them .

To get the implication on how large is the BGG brightening for these $\Nrich < 2$ groups,
one way is to go to a sample of groups with deeper magnitude limit. With a lower
magnitude limit to the member galaxies, some of the $\Nrich < 2$ groups become to have
more than two members and so that the OS studies can be applied. We provide such an
analysis in Appendix A, where we select a sample of  lower redshift groups with members
complete to $\Mr < -18.5$ mag.  By including  fainter members, we show that the BGGs of
$\Nrich < 2$  groups also have experienced a significant brightening process, i.e. $\Delta M
> 0$. Moreover, we further show that the brightening process of the BGGs of the low mass
groups varies  among different groups and shows  complicate dependence on the group
richness parameters. The physics behind this variety is that, the BGG growth history depends
on  its real halo mass and environment, while for the low mass halos, any single observable
parameter is not adequate to characterize its real halo mass and/or  environment. We will
leave a detailed discussion of the dependence of $\Delta M$ on the group properties in a
forthcoming paper.

\subsection{The average stellar mass increment of the brightest group
galaxies}\label{Sec_Mass}

The systematic brightening of BGGs may be caused either by extra star formation (which
makes them bluer and brighter) or by an increase in stellar mass. In order to partly
distinguish these two possibilities we also carry out modeling in terms of the stellar masses
of the BGGs. In the group catalog used here, the stellar mass ($M_{\rm s}$) of each group
member galaxy was obtained from the relation between the stellar mass-to-light ratio and
the color \citep{Bell03}.  We thus make a similar study on the mass distribution of the most
massive galaxies (MMGs) as we have done for the BGGs by assuming that all MMGs have
increased their stellar masses on average  by amounts $\Delta \Log M_{\rm s}$ in logarithm
unit. Since the analysis is parallel with that for BGGs, the details are given in Appendix B.

\begin{figure}
 \includegraphics[width=85mm]{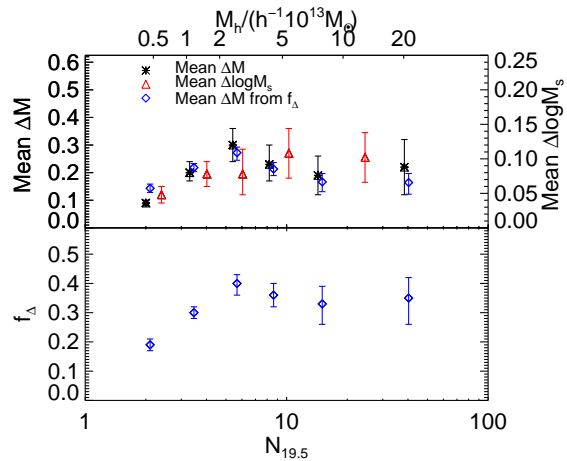}
\caption{The best estimates of the average magnitude brightening $\Delta
\bar{M}$ of the BGGs  as functions of the $\Nrich$ and host   halo mass (top panel). Stars represent
$\Delta\bar{M}$ estimated from the constant BGG   brightening model
(Section \ref{Sec_Cmodel}), while the diamonds show that estimated
from the gap dependent BGG brightening model
(Section   \ref{Sec_Gmodel}). The triangles show the  average stellar mass increment $\Delta \Log \bar{M}_s$ of the MMGs
(labelled on the   right vertical axis)  (Section \ref{Sec_Mass}). Bottom: The best
estimate of $f_\Delta$ in the gap-dependent BGG
brightening model (Section  \ref{Sec_Gmodel}).}
\label{Fig_DeltaM}
\end{figure}

Here we make use of the groups whose MMGs are also the BGGs (99\% of
  the cases), and plot the best estimate of the average mass increment
$\Delta\Log \bar{M}_{\rm   s}$ as a function of group halo mass $\Mh$ in the top panel of
Fig. \ref{Fig_DeltaM}. The error-bars also represent the 68 percentiles of the likelihood
distribution at each side of  the best estimate. Note that  the ranges of the left vertical axis
($\Delta \bar{M}$ ) and the right vertical axis ($\Delta \Log \bar{M}_{\rm s}$) of the top
panel of Fig. \ref{Fig_DeltaM} are $0.0-0.625$ mag and $0.0-0.25$ dex, respectively, and so
the results show that the values of $\Delta\bar{M}$ and $\Delta \Log \bar{M}_{\rm s}$ are
consistent, suggesting that the brightening of BGGs is mainly caused by the increase in
stellar mass. The amount of average mass increment is about twenty percent, $\Delta\Log
\bar{M}_s\sim 0.09$, for  the massive haloes with $\Mh>10^{13}\Msun$.  For the low mass
haloes in the lowest richness bin,  the stellar mass increment is also lower, about 10 percent
($\Delta\Log \bar{M}_s\sim 0.05$). Again, we caution that our low mass groups are not
complete when in terms of halo mass.

\subsection{The magnitude gap $G_{1,2}$ distribution}\label{Sec_gap}

So far we have studied on average how much  BGGs are  brightened by modeling the
absolute magnitude distributions of  the BGGs and SGGs together. In this subsection we
study how the brightening is correlated with the other group properties. e.g. the magnitude
gap $G_{1,2}$.  The distribution of $G_{1,2}$ and its correlation with the host halo mass have
been explored by many recent studies\citep{More12,Paranjape12,Tal12}. Some studies find
that the $G_{1,2}$ distribution may contain dynamical information of the groups that is not
contained in the group richness and BGG luminosity \citep{Hearin13b,Hearin13a}.

If we have a model that describes how BGG is brightened in individual groups, we will be
able to explain not only the $M_1$ and $M_2$ distributions but also the distribution of the
magnitude gap $G_{1,2}$. However, we find that the $G^*_{1,2} \equiv (M_2 - M^*_1)$
distribution can not be reproduced by our best constant BGG brightening model. For the
best estimates of $\Delta \bar{M}$ in Section \ref{Sec_Cmodel}, the KS test probability
$P_{\rm GAP}$(the  $G^*_{1,2} $ distribution follows the same distribution as the shuffled
groups) is nearly zero ($<10^{-5}$) for all richness bins. Moreover, except the two highest
$\Nrich$ bins that have relative low statistical significance, we can not find any values of
$\Delta M$ in the constant BGG brightening model which could make $P_{\rm GAP} > 0.01$
(see the dot-dashed lines in Fig. \ref{Fig_modelBS}). This result implies that our simple
constant BGG brightening model is too simplistic to account for individual brightening. The
inability of the constant brightening model in reproducing the $G^*_{1,2}$ distribution can
be understood in another way. If we assume that the initial $G^*_{1,2}$ distribution  follows
the OS, it will always peak at $G^*_{1,2}=0$ for any Schechter-like CLFs of group members
\citep[e.g.][]{Paranjape12}.  If all initial BGGs are then brightened by an amount of
$\Delta\bar{M}$ in the subsequent evolution, the final $G_{1,2}$ distribution will peak at
$\Delta\bar{M}$, in contrast to the observed $G_{1,2}$ distribution, which still peaks around
0(see bottom right panel of Fig. \ref{Fig_Basic}).

\subsection{Magnitude gap-dependent brightening of the brightest
  groups galaxies} \label{Sec_Gmodel}

As a simple gap-dependent model, we assume that the brightening
magnitude $\Delta M$ of each BGG is proportional to the magnitude
gap $G_{1,2}$ of its host group,
\begin{equation}\label{Equ_dM}
\Delta M=f_\Delta*G_{1,2}
\end{equation}
where the coefficient $f_\Delta$ is a model parameter to be
determined\footnote{Theoretically,  it would be better to relate the brightening with the
initial magnitude gap $G^*_{1,2}$. In practice, however, we associate it with the
observational magnitude gap, $G_{1,2}$, so that we do not need to make iterations. }.  To
distinguish this model from the constant brightening model in Section \ref{Sec_Cmodel}, we
refer it as the $G$ model in the following.

In general, if a group has a larger initial gap $G^*_{1,2}$, its BGG is more dominant within
the group.  The more dominant BGG potentially has the ability to grow more through e.g.
galaxy cannibalism.  Thus,  the observed $G_{1,2}$ becomes even larger.  With the
parametrization of Equ. \ref{Equ_dM}, the value of $f_\Delta$ lies in between 0 and 1. The
limit $f_\Delta=0$ is the case that there is no BGG brightening at all. On the other hand, the
limit $f_\Delta=1$ means that the observed magnitude gap $G_{1,2}$ is totally contributed
by the BGG brightening process.

As in Section \ref{Sec_Cmodel}, we apply a gap dependent dimming model to the BGGs in
each $\Nrich$ bin. For any given values of $f_\Delta$,  the BGGs are dimmed by $\Delta M$
using Equ. \ref{Equ_dM}. We then make OS predictions for the $M^*_1$, $M_2$ and
$G^*_{1,2}$ distributions by shuffling group members as that done in Section
\ref{Sec_Cmodel}. We explore the parameter space of $f_\Delta$ by calculating the
combined KS probability,
\begin{equation}
 P_{\rm B,S,G}=P_{\rm BGG}*P_{\rm SGG}*P_{\rm GAP}\,,
\end{equation}
where $P_{\rm BGG}$, $P_{\rm SGG}$ and $P_{\rm GAP}$ are the KS probabilities that the
distributions of the  BGG magnitudes, SGG magnitudes and magnitude gaps of the
BGG-dimmed sample follow the OS predictions respectively. Compared with $P_{\rm
B,S}$(Equ. \ref{Equ_BS}) used for the $C$ model in Section \ref{Sec_Cmodel}, the new model
uses the  $G^*_{1,2}$ distribution as an extra constraint.

\begin{figure}
\includegraphics[width=85mm]{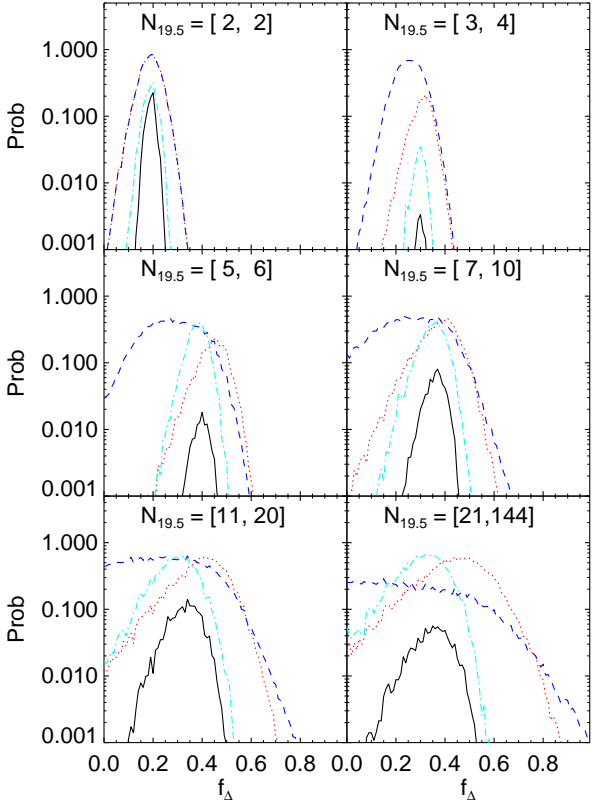}
\caption{The KS test  probabilities  as function of $f_\Delta$  for  6
  different $\Nrich$ bins. The dotted, dashed and dot-dashed lines
  show the probabilities$P_{\rm BGG}, P_{\rm SGG}, P_{\rm GAP}$ respectively, while
the  solid lines show the combined   probability $P_{\rm B,S,G}$ (see Section \ref{Sec_Gmodel} for detail) }
\label{Fig_modelall}
\end{figure}

We show the  $P_{\rm B,S,G}$  as function of  $f_\Delta$ with solid lines for different
$\Nrich$ bins in Fig. \ref{Fig_modelall}, whereas the $P_{\rm BGG}$, $P_{\rm SGG}$ and
$P_{\rm GAP}$ are shown by the dotted, dashed and dot-dashed lines respectively.
Remarkably, the constraints from the BGGs($P_{\rm BGG}$), SGGs($P_{\rm SGG})$ and the
magnitude gap($P_{\rm GAP}$) on this gap dependent BGG brightening model are in good
consistence with each other. As in Section \ref{Sec_Cmodel} , we take the value of $f_\Delta$,
where the accumulated $P_{\rm B,S,G}$ from $f_\Delta=0$ is half of the accumulated
$P_{\rm B,S,G}$ from $f_\Delta=0$ to $f_\Delta=1$, as the best model estimate.  The 68
percentiles on each side of the best model are then used as the error estimates. The best
estimates of $f_\Delta$  are about $0.2$-$0.4$, with weak dependence on $\Nrich$. They
are plotted  in the bottom panel of Fig. \ref{Fig_DeltaM} and  listed in Table 1.

For each richness bin, we also calculate the average BGG dimming/brightening from Equ.
\ref{Equ_dM} using the best estimates of $f_\Delta$ and plot them as the open diamonds in
the top panel of Fig. \ref{Fig_DeltaM}, where the error-bars are obtained from the error
estimates of $f_\Delta$. As we can see, the average BGG dimming/brightening of the $G$
model is in excellent agreement with that of the  $C$ model in Section \ref{Sec_Cmodel}.
Moreover, because of the further constraint from the $G^*_{1,2}$ distribution, the
constraints on the average BGG brightening in the $G$ model are tighter than that in the
$C$ model .

\begin{figure}
\includegraphics[width=85mm]{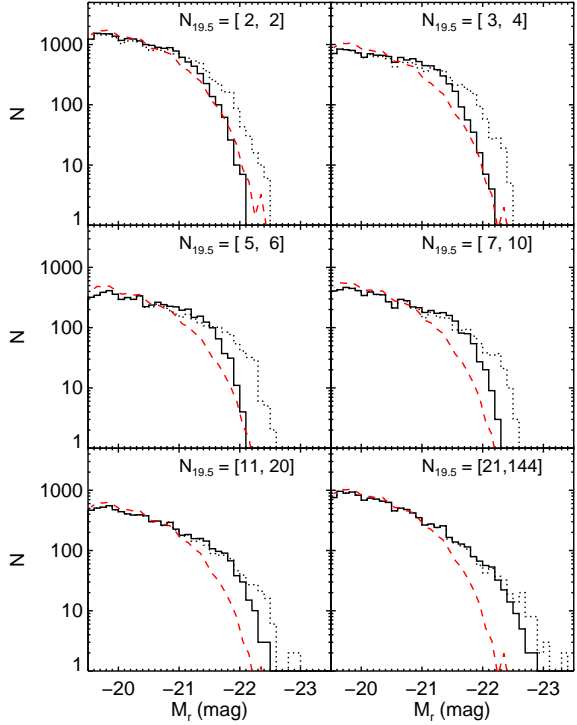}
\caption{The primordial  conditional luminosity function of the group members
in six $\Nrich$ bins (solid histograms). In each panel, the dotted
histogram show the observed conditional luminosity function, while the dashed line
show the  luminosity function of the single galaxies which has been normalized to the
number of the group members in that richness bin. }
\label{Fig_OCLF}
\end{figure}

To have a better visualization of our gap dependent BGG brightening model, we make
predictions for the $M_1$, $M_2$ and $G_{1,2}$ distributions as functions of $\Nrich$, and
compare them with the corresponding observational results. To do this, we  first construct a
`BGG-dimmed' sample using the best model of $f_\Delta$  for each $\Nrich$ bin.  We then
use the $\Mr$ distributions of the `BGG-dimmed' sample as the primordial CLFs of the
group members. These primordial CLFs are shown as the solid histograms in Fig.
\ref{Fig_OCLF}, where the observed CLFs (already shown as the solid histograms in Fig.
\ref{Fig_CLF}) are shown as the dotted histograms for comparison.  For reference, we also
plot the luminosity distribution of the single galaxies as the dashed curve in each panel after
normalized to the same number of the group members in the corresponding $\Nrich$ bin.
By comparing  with the luminosity distribution of the single galaxies, we see that, after
considering the BGG brightening in different $\Nrich$ bins,  the primordial CLF still changes
systematically with group richness , in the sense that group members are also on average
brighter in groups with higher richness.

\begin{figure}
\includegraphics[width=85mm]{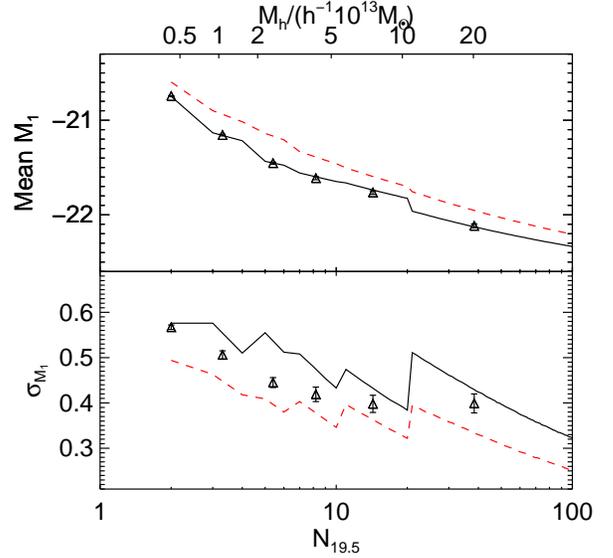}
\caption{The  mean and scatter of the magnitude of BGGs ($M_1$) as function of
$\Nrich$ and $\Mh$. The abscissa of  $\Mh$ is labelled on top of the figure.
The triangles show the results of our real groups. The
dashed lines show the model prediction from OS, while the solid lines
show the results after applying a gap-dependent  BGG brightening.  }
\label{mod_BGG}
\end{figure}

\begin{figure}
\includegraphics[width=85mm]{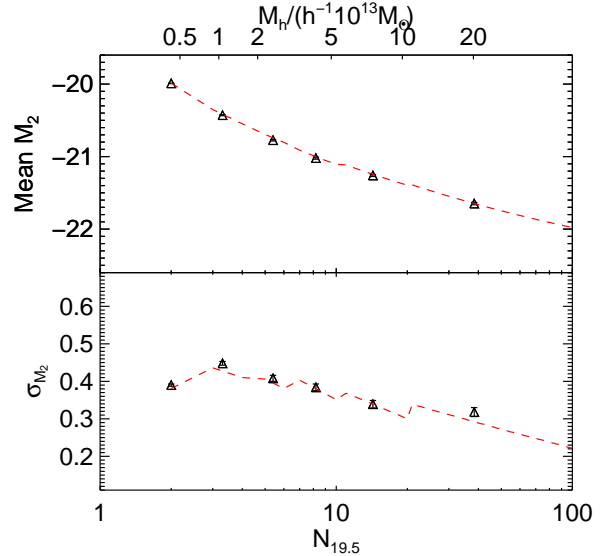}
\caption{The  mean and scatter of the magnitude of SGGs $M_2$ as function of
$\Nrich$ and $\Mh$. The triangles show the results of our real groups, while the
dashed lines show our model prediction from OS.  }
\label{mod_SGG}
\end{figure}

\begin{figure}
\includegraphics[width=85mm]{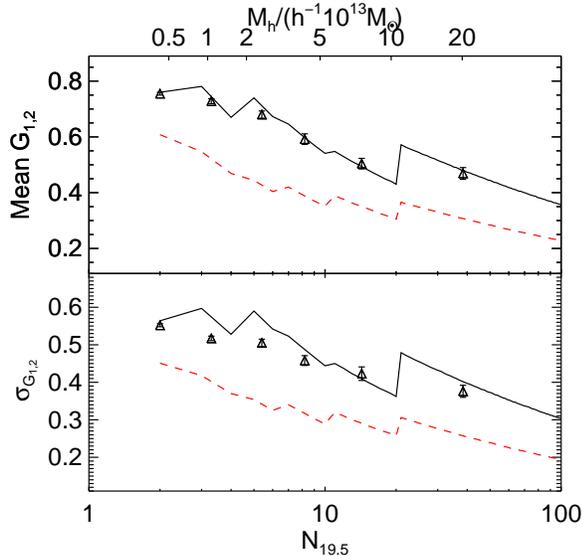}
\caption{The  mean and scatter of the magnitude gap $G_{1,2}$ as function of
$\Nrich$ and $\Mh$. The triangles show the results of our real groups. The
dashed lines show the model prediction from OS, while the solid line
shows the results after applying a gap-dependent  BGG brightening.  }
\label{mod_GAP}
\end{figure}

With the  primordial  CLFs in different $\Nrich$ bins, we generate 100,000 mock groups for
each $\Nrich$ using a Monte-Carlo method, and obtain $M^*_1$ and $M_2$ for each mock
group by sorting the group members in luminosity ranks. Next, we brighten $M^*_1$ of
each mock group with an amount of
\begin{equation}\label{Equ_dM2}
\Delta M =f_\Delta/(1-f_\Delta)*G^*_{1,2} \,,
\end{equation}
where $f_\Delta$  takes the best estimate shown in Fig. \ref{Fig_DeltaM}.  In Equ.
\ref{Equ_dM2}, $G^*_{1,2} $ is the magnitude gap of the OS sample, i.e. the gap before the
BGG brightening, so that  Equ. \ref{Equ_dM2}  takes a different formulism from Equ.
\ref{Equ_dM}. These new mock groups after BGG brightening are referred to as the model
groups, and have $M_1=M^*_1-\Delta M$ and $G_{1,2}=G^*_{1,2}+\Delta M$. We then
calculate the mean and dispersion of the $M_1$, $M_2$ and $G_{1,2}$ distributions of the
model groups for each $\Nrich$. The mean and dispersion so obtained as functions of
$\Nrich$ and $\Mh$ are plotted as the solid lines in Fig. \ref{mod_BGG},
\ref{mod_SGG},\ref{mod_GAP}, respectively. Eq. (\ref{Equ_NMh}) is used in converting
$\Nrich$ to $\Mh$. Since the number of model groups for each $\Nrich$ is large, their
statistical fluctuation is small. For comparison, the results of the real groups are plotted as
the triangles in these figures.  In Fig. \ref{mod_BGG} and \ref{mod_GAP}, we also plot the
results of the model groups before the BGG brightening (i.e. OS model predictions) as the
dashed lines. As one can see, after applying a simple BGG brightening, the distributions of
$M_1$ and $G_{1,2}$ both become in good agreement with the observed distributions. For
the SGGs shown in Fig. \ref{mod_SGG}, there are excellent agreements between the
observations and OS predictions. Similar conclusions have already been shown in Fig.
\ref{OS_SGG}. However, these two plots have different implications. Fig. \ref{OS_SGG} shows
that the SGGs are consistent with the extreme value distribution of the satellite population
and are independent of the BGGs. Here in Fig.\ref{mod_SGG}, on the other hand, the SGGs
and BGGs are assumed to originate from the same population, and  the SGGs are in good
consistence with OS predictions after the BGG brightening has been taken into account.

\section{The Tremaine-Richstone test}\label{Sec_TR}

\citet{TR77} defined two parameters to test the statistical nature of BGGs (hereafter TR
tests) using the dispersion of the absolute magnitude of the BGGs, $\sigma_{\rm M_1}$, and
the mean ($\bar{G}_{1,2}$) and dispersion ($\sigma_{\rm G_{1,2}}$) of the magnitude gap
$G_{1,2}$:
\begin{eqnarray}\label{Equ_TR}
\nonumber
   T_1 \equiv \frac{\sigma_{\rm M_1}}{\bar{G}_{1,2}}\,; \\
    T_2 \equiv   \frac{\sigma_{\rm G_{1,2}}}{\bar{G}_{1,2}}\,.
\end{eqnarray}
For the tail-end of a statistical distribution with an exponential decay, \cite{TR77} showed
that these two test parameters have $T_1 \geq 1$ and $T_2 \geq 0.82$.  That means, if the
BGGs are drawn from the same population as the other members, the expected magnitude
gap $\bar{G}_{1,2}$  should  be smaller than the scatter of the BGG magnitudes and  be of
the same order as its own dispersion.  The TR tests are not sensitive to the shape of the
assumed luminosity function (except the exponential decay) and the variation from cluster
to cluster, so are widely used to test whether the brightest cluster galaxies are consistent
with an extreme value population \citep{Geller83, Postman95, Loh06, Lin10}. However, the
applicability of the TR tests has not  been validated for the low mass/richness groups.

\begin{figure}
\centering
\includegraphics[width=85mm]{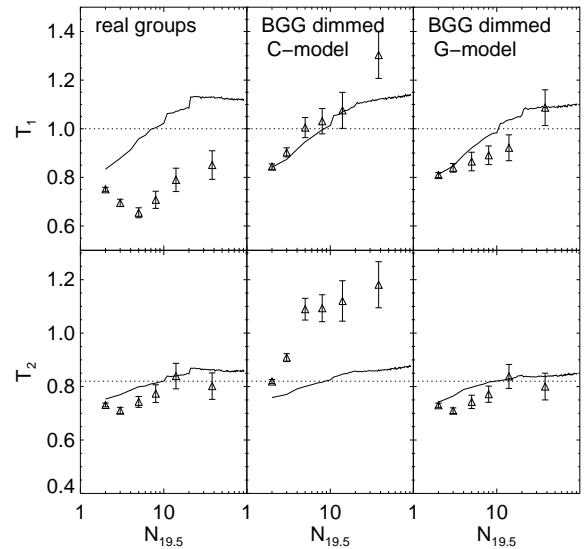}
\caption{The TR-test parameters as function of group richness.  The left,
  middle and right panels show the results for different group samples.  In
  each panel, the triangles show the TR parameters of the group sample in
  different richness bins, while the solid line shows the statistical model
  prediction from Monte-Carlo simulations.  The horizontal lines show the
  critical values of \citet{TR77}.  Left: real groups; Middle: groups
  whose BGGs have been dimmed by the best $C$  model (Section
  \ref{Sec_Cmodel}); Right: groups whose BGGs have been dimmed by the best
  $G$ model (Section \ref{Sec_Gmodel}).} \label{Fig_TR}
\end{figure}

We carry out the TR tests for three different group samples. The first is the real group
sample. The second is the group sample with the BGGs {\it dimmed} using the best $C$
model (Section \ref{Sec_Cmodel}), while the third one is the group sample whose BGGs are
{\it dimmed} using the best $G$ model (Section \ref{Sec_Gmodel}).  All three group samples
are also divided into six $\Nrich$ bins as before.  The TR test values are calculated for
groups within each $\Nrich$ bin, and the results are shown as the triangles in Fig.
\ref{Fig_TR}, with the left, middle and right panels showing the results for the three different
samples, respectively.  The error-bars of the TR-tests are estimated from 100  boot-strap
re-samplings.

For each sample, we also calculate the values of TR parameters predicted by the OS as
functions of group richness with Monte-Carlo simulations. Specifically, we generate 100,000
random groups for each $\Nrich$, as is done in Section \ref{Sec_Gmodel}. Note that the
underlying distributions of these three group samples are different and  our OS predictions
are made in a self-consistent way. The values of the TR parameters predicted by the OS as
functions of $\Nrich$ are shown as the solid line in each panel of Fig. \ref{Fig_TR}. Again,
because of the large size of the Monte-Carlo sample, the uncertainties of the predicted
values are negligible.  The critical TR test values suggested for rich clusters by \citet{TR77} ,
$T_1=1$, and $T_2 = 0.82$ are shown as the dotted horizontal lines in each panel.

As one can see from the solid lines in different panels of Fig. \ref{Fig_TR}, the dependence of
the OS predictions on the underlying distributions is quite weak. This result confirms that
the TR test is not sensitive to the shape of the luminosity function of the group members.
However, there is a weak dependence of the test parameters on the group richness. Both
$T_1$ and $T_2$ increase monotonically but slowly with richness and approach their
asymptotic values at $\Nrich > 30$. The two criteria, $T_1 \geq 1$ and $T_2 \geq 0.82$, are
achieved only at $\Nrich>10$, while the expectation values $T_1=1.29$ and $T_2=1$
suggested by \citet{Bhavsar85} are not achieved even at $\Nrich=100$.

For the real groups, $T_1$ is systematically smaller than the statistical model prediction,
while $T_2$ shows agreement with the model (left panels of Fig. \ref{Fig_TR}). Because BGGs
are expected to have experienced brightening, the mean magnitude gap $\bar{G}_{1,2}$ of
the real groups are systematically larger than the OS prediction (Fig. \ref{mod_GAP} ). As a
result, $T_1$ is systematically smaller than the OS prediction. On the other hand, since the
scatter in the $G_{1,2}$ of the real groups is also systematically larger than the OS
prediction (Fig. \ref{mod_GAP}), the test parameter $T_2$ appears consistent with the OS
prediction. Because of this coincidence, different answers can be obtained by using different
TR tests for high-richness groups. This has been noticed by \citet{Lin10}, who found that the
BGGs of high-mass clusters are not consistent with a statistical sample based on $T_1$, but
are consistent when $T_2$ is used.

For the two different BGG-dimmed samples, the TR test values are quite different. For the
constant BGG dimming model (middle panels, $C$ model in Section \ref{Sec_Cmodel}),
$T_1$ is consistent with the OS prediction, while $T_2$ becomes systematically larger than
the OS prediction. For the gap-dependent BGG dimming model (right panels, $G$ model in
Section \ref{Sec_Gmodel}), {\it both} of the test parameters are roughly consistent with the
OS predictions.  The different behaviors of the two BGG-dimmed models are mainly caused
by the way how the BGGs are dimmed. For the $C$ model, the BGGs are dimmed by a
richness-dependent constant, the scatter in $G_{1,2}$ is thus roughly preserved, so that the
value of $T_2$ increases systematically due to the BBG dimming. For the $G$ model, on the
other hand, the mean and scatter of $G_{1,2}$ are both decreased due to the
gap-dependent dimming, so that the values of $T_2$ remain unchanged after BGG
dimming. This can be understood in another way, the $T_2$ values of primordial groups
before the BGG brightening are expected to be consistent with the OS prediction. Thus, only
when the BGG brightening process changes the mean and scatter of $G_{1,2}$ in a similar
manner, can the $T_2$ values of the evolved groups remain consistent with the OS
prediction. In this sense, the agreement shown in the lower left panel of Fig. \ref{Fig_TR} may
not be a coincidence, but is a result of how BGGs are brightened.

Given these different behaviors of $T_1$ and $T_2$,  it seems that $T_1$ is more useful  than
$T_2$ in testing whether BGGs are consistent with the OS, while $T_2$ may be useful in
testing the details how the BGGs are brightened.

\section{Summary}\label{Sec_sum}

In this study, we test the statistical hypothesis of BGGs using a large sample of groups of
galaxies selected from the DR7 of SDSS. We define a richness parameter $\Nrich$, the
number of the group members with $\Mr<-19.5$ mag, to parameterize the host halo mass
of each group. By dividing the groups into different $\Nrich$ bins, we measure the
conditional luminosity distribution of the group members and build the corresponding
statistical sample of groups by Monte-Carlo simulations. By comparing the real groups and
the statistical sample, we study the statistical properties of the BGGs and other members.
Our results can be summarized as follows:
\begin{itemize}
\item The BGGs are systemically brighter than the OS model prediction, which can
  be explained by a special BGG brightening process.
\item To make the distribution of the BGG luminosities consistent with the EVS,
  the BGGs on average have to be dimmed by about $\sim 0.2$ mag.
\item Taking into account the BGG brightening, the luminosity
  distribution of the SGGs is in good agreement with the OS.
\item To simultaneously reproduce the distributions of the magnitudes of BGG, SGG and
    their gap, the brightening of a BGG should be correlated with the
    gap $G_{1,2}$ (or $G^*_{1,2}$). Such a brightening will boost
   $G^*_{1,2}$ by an amount of $\sim 50$ percent relative to the EVS
   of the underlying distribution.
\end{itemize}

The above results may provide some insight into the formation and evolution of the BGGs.
The BGGs may have originated  from  the same statistical processes as other bright galaxies.
On top of that,  BGGs should also have experienced some brightening process which not be
experienced by other bright galaxies. The brightening is mainly due to the growth of stellar
mass, and may be related to local processes, such as galactic cannibalism and/or star
formation in cooling flows. On average, the excess over the expectation of the EVS is about
20 percent for groups with halo masses $\Mh > 10^{13} \Msun$. The excess of stellar mass
of a BGG is more significant if itself is more prominent inside its group.  Such a scenario is in
good agreement with the evolution of BGGs through minor mergers
\citep[e.g.][]{Edwards12}.

In such a scenario, the BGGs are the central galaxies in groups. However, as shown by
\citet{Skibba11}, the BGGs may not always be the centrals.  The fraction of non-central BGGs
may be as high as 25 to 40 percent. We have tested such a possibility by assuming that a
fraction of BGGs, $f_{\rm BNC}$, are not the centrals and so have not experienced any
brightening effects (i.e. $\Delta M$=0). We find such  a scenario is also acceptable,  but our
statistical model cannot provide any strong constraint on $f_{\rm BNC}$ could be. In the test
we have also assumed that the current BGGs are the initial BGGs. It is also possible that an
initial central galaxy is not the BGG but  becomes the BGG only after the brightening
process. For the $C$ model, our statistical data  cannot provide any constraints on such a
possibility, as discussed in Section \ref{Sec_Cmodel}. Such a scenario is not compatible with
the $G$ model by definition. Finally, we emphasize again that our results are statistical. How
much the BGGs are brightened for specific groups may depend on the details of their
formation processes(as we haven shown in Appendix A). In a future paper, we will divide
groups into subsamples according to intrinsic properties in addition to group richness, and
examine how the growth of the BGGs may depend on these other properties.

\section*{Acknowledgments}
We thank the anonymous  referee for helpful comments which significantly clarified the text.
This work is supported by the grants from NSFC (Nos. ,10878003, 10925314, 11128306,
11121062, 11233005), 973 Program 2014CB845705, the Shanghai Municipal Science and
Technology Commission No. 04dz\_05905 and CAS/SAFEA International Partnership
Program for Creative Research Teams (KJCX2-YW-T23).  HJM would like to acknowledge the
support of NSF AST-1109354 and NSF AST-0908334.

\clearpage

\begin{longtable*}{lcccccc}
\caption{The average halo mass $\Mh$, the number of groups $N_{\rm {gro}}$,
  the number of group members $N_{\rm{gal}}$, the  best estimations of model parameters
   $\Delta\bar{M}$(Section \ref{Sec_Cmodel}), $f_\Delta$(Section \ref{Sec_Gmodel}) in six $\Nrich$ bins. }
\endhead
   \hline   \\
  $\Nrich$ & 2-2 & 3-4 & 5-6 & 7-10 & 11-20 & 21-144 \\
  \hline
  $N_{\rm{gro}}$ & 9,817 & 3,620 & 1,013 & 765 & 494 & 305 \\
  $\Log\,\Mh$ & 12.61 &  13.05 &  13.41 & 13.63 & 13.88 &  14.30 \\
  $N_{\rm{gal}}$ & 19,634 & 11,958 & 5,457 & 6,261 & 7,072 & 11,699\\
$\Delta\bar{M}$ & $0.09^{+0.01}_{-0.01}$ & $0.20^{+0.03}_{-0.04}$ &
$0.30^{+0.06}_{-0.06}$ & $0.23^{+0.07}_{-0.06}$ & $0.19^{+0.07}_{-0.07}$ & $0.22^{+0.10}_{-0.10} $\\
 $f_\Delta$ & $0.19^{+0.02}_{-0.02}$ & $0.30^{+0.02}_{-0.02}$ & $0.40^{+0.03}_{-0.04}$ &
$0.36^{+0.04}_{-0.04}$ & $0.33^{+0.06}_{-0.07}$ & $0.35^{+0.09}_{-0.09}$\\
\hline
\end{longtable*}

\appendix

\section{The group richness $N_{18.5}$}

In this section, we test the BGG brightening model with a new sample of groups whose
complete magnitude limit reaches to $-18.5$ mag by applying a  lower redshift limit
($z<0.506$) to the group catalog of \citet{Yang07}.  Similar as $\Nrich$,   $N_{18.5}$ is
defined as  the number of the group members brighter than $-18.5$ mag. In the DR7
version of the group catalog of \citet{Yang07},   there are 54,497 groups with $N_{18.5} \ge
1$ at $z<0.506$. The number of their members is 80,761. We name this new group sample
as the $N_{18.5}$ groups so as to be distinguished from the $\Nrich$ groups we used in the
main text. Because of the including of the fainter members, the richest group now has
$N_{18.5}=284$.

\begin{figure}
\centering
\includegraphics[width=85mm]{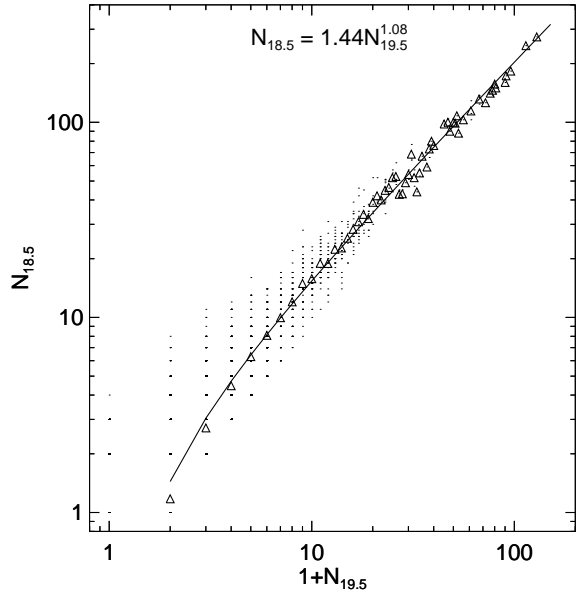}
\caption{The correlation between  $N_{18.5}$ and $\Nrich$ for the $N_{18.5}$ groups . The triangles
show the mean $N_{18.5}$ at each given $\Nrich$. The solid line shows the fitting line of Equ. \ref{Equ_N2}}.
\label{Fig_N185}
\end{figure}

For the $N_{18.5}$ groups, we calculate both $N_{18.5}$ and $\Nrich$ for each group. In
Fig. \ref{Fig_N185}, we plot their $\Nrich$ versus $N_{18.5}$ . At each $\Nrich$ , we
calculate the mean $N_{18.5}$ and plot them as the triangles. As we can see, there is a tight
correlation between$N_{18.5}$ and $\Nrich$ for high richness groups. For the low richness
groups, there are  significant scatters between each other.  For the $\Nrich$ groups in the
main text, all of them have $\Nrich \ge 1$ .  However, for the $N_{18.5}$ groups, there are a
significant fraction of groups with $\Nrich=0$ . These are the groups with all their members
fainter than $-19.5$ mag but with at least one member brighter than $-18.5$ mag. The
number of such $\Nrich=0$ groups is as high as 26,625 and their $N_{18.5}$ ranges from 1
to 4. Also, because the $\Nrich=0$ groups have no member brighter than $-19.5$ mag, no
halo mass has been estimated in the group catalog of \citet{Yang07}.

We  fit  a power law relation between $N_{18.5}$ and $\Nrich$ for $\Nrich > 0$ groups and
obtain
\begin{equation}\label{Equ_N2}
N_{18.5}=1.44N_{19.5}^{1.08}\,.
\end{equation}
This fitting relation is shown as the solid line in Fig. \ref{Fig_N185}. As we can see, this
relation fits the mean $N_{18.5}$  as function of  $\Nrich$ very well.

\begin{figure}
\centering
\includegraphics[width=85mm]{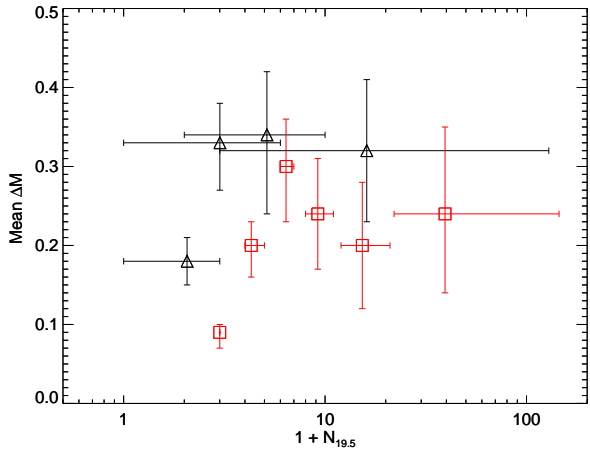}
\caption{The average BGG brightening($\Delta \bar{M}$) of the groups as function of  $\Nrich$.
 The triangles show $\Delta \bar{M}$ of the groups in 4 $N_{18.5}$ bins,  while the squares show the groups in 6
 $\Nrich$ bins (see  Fig. \ref{Fig_DeltaM}). The vertical error-bars show the 68 percentiles of the likelihood
distribution, while the horizontal error-bars represent the $\Nrich$ ranges  of
 the groups in each given bin.}
\label{Fig_DM185}
\end{figure}

To  test our BGG brightening model with the new $N_{18.5}$ sample, we make analysis on
the average BGG  brightening($\Delta \bar{M}$) using the same algorithm as that in Section
\ref{Sec_Cmodel}.We bin the  $N_{18.5}$ groups into 4 $N_{18.5}$ bins( [2,2], [3,5], [6,10],
[11,274]), where the numbers of the groups are 5069, 2148, 519 and 441 respectively. The
best estimates $\Delta \bar{M}$ for the groups in the four bins are 0.18, 0.34, 0.33 and 0.32
mag respectively. To compare with the results of $\Nrich$ groups in Section
\ref{Sec_Cmodel}, we show the $\Delta \bar{M}$ as function of their average $\Nrich$ for
 the $N_{18.5}$ groups  as the triangles in Fig. \ref{Fig_DM185}.  The  results of $\Nrich$ groups in Section
\ref{Sec_Cmodel} so that are plotted as the open squares for comparison.

As we can see, for the high richness groups, the $\Delta \bar{M}$ from the $N_{18.5}$
groups are in consistence with the $\Nrich$  groups. However, for the low richness groups,
the results show discrepancy. The $\Delta \bar{M}$ of the $\Nrich=2$ groups is $\sim 0.1$
mag, whereas the $\Delta \bar{M}$ of  the groups in the $N_{18.5}=[3,5]$ bin  is $\sim 0.3$
mag although their average  $\Nrich$ is also about 2.   This discrepancy is caused by the
fact that there are significant scatters between $N_{18.5}$ and $\Nrich$ for the low richness
groups. For example, although the mean $\Nrich$ of the groups in the $N_{18.5}=[3,5]$ bin
is close to 2, their $\Nrich$ ranges are from 0 to 5.  There are 2148 groups within the
$N_{18.5}=[3,5]$ bin, however, only less than half  of of them(893) have $\Nrich=2$.

Besides the large scatter between $\Nrich$ and $N_{18.5}$, the large discrepancy in $\Delta
\bar{M}$ for the low richness group may also imply that the growth of the BGGs may not
depend on any single richness parameter ($N_{18.5}$ or $\Nrich$) in a simple way. To show
this, we make an analysis on the dependence of $\Delta M$ on both $N_{18.5}$ and
$\Nrich$. In the $N_{18.5}$ group sample,  there are 2,225 groups with $\Nrich=2$ and
their $N_{18.5}$ range from 2 to 11. For these $\Nrich=2$ groups, a group with $N_{18.5} =
2$ means that their is no group member with luminosity between $-19.5$ and $-18.5$
mag. On the other hand, a group with $N_{18.5} = 11$ means it has 9 members within the
luminosity range $-19.5 < M_r \le -18.5$. Therefore, our test is whether the brightening of
the BGGs is correlated with the number of the group members in between $-19.5 < M_r \le
-18.5$.

\begin{figure}
\centering
\includegraphics[width=85mm]{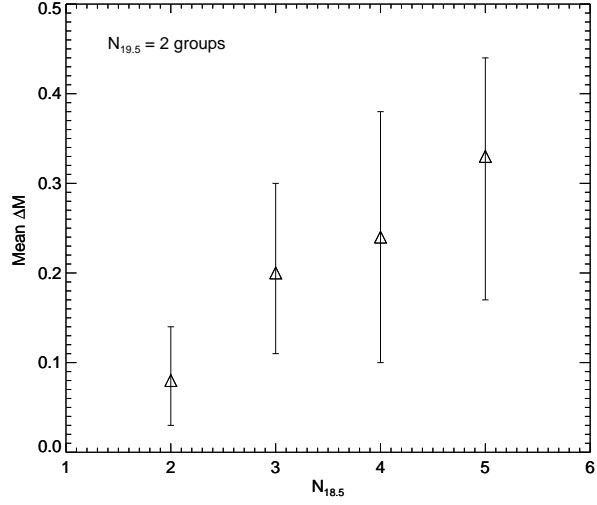}
\caption{The average BGG brightening($\Delta \bar{M}$) of the $\Nrich=2$  groups as function of their $N_{18.5}$.}
\label{Fig_N2_185}
\end{figure}

 We divide the above 2,225 $\Nrich=2$ groups into four sub-samples according to their $N_{18.5}$ values.
The four sub-samples are the groups with  $N_{18.5}$  being 2,3,4 and $>4$  respectively.
The number of the groups in these four sub-sample are 1270, 573, 232 and 150 accordingly.
We estimate the $\Delta \bar{M}$ of the BGGs for these 4 $\Nrich=2$ sub-samples using
the same algorithm as that in Section \ref{Sec_Cmodel}.  The best estimates of the  $\Delta
\bar{M}$ of these 4 sub-samples are shown in Fig. \ref{Fig_N2_185}. As can be seen,
although all the sub-samples have $\Nrich=2$, the $\Delta \bar{M}$ shows a strong
dependence on $N_{18.5}$. The $\Delta \bar{M}$ is systematically larger for the groups with
larger $N_{18.5}$. The sub-sample of  the groups with $N_{18.5}=2$ is the majority of the
global $\Nrich =2$ sample(1270 of 2225), which might be the reason that the $\Delta M$
of this sub-sample is close to the result of the general $\Nrich =2$ sample(Section
\ref{Sec_Cmodel}, $\sim 0.1$ mag).  The increase of $\Delta \bar{M}$ with the increasing
number of faint members($-19.5 < M_r \le -18.5$)  might be qualitatively explained by that
the BGGs can grow more through cannibalization if they have more faint satellites. However,
a quantitative explanation needs a more detailed modeling of the growth of the BGGs.  We
will leave such a detailed study in forthcoming.

\section{The most massive galaxies of groups}\label{MMG}

As described in \cite{Yang07}, the stellar mass of all group member galaxies
have been computed using fitting formula of \citet{Bell03},
\begin{equation}
\Log [M_{\rm s}/(h^{-2}M_\odot)]=-0.406+1.097[^{0.0}(g-r)]-0.4(^{0.0}\Mr-5\Log
h-4.64)\,,
\end{equation}
where $^{0.0}(g-r)$ and $^{0.0}\Mr$ are the SDSS $g-r$ color and $r$ band absolute
magnitude $K$-corrected and evolution corrected to redshift $z=0.0$.

\begin{figure}
\centering
\includegraphics[width=100mm]{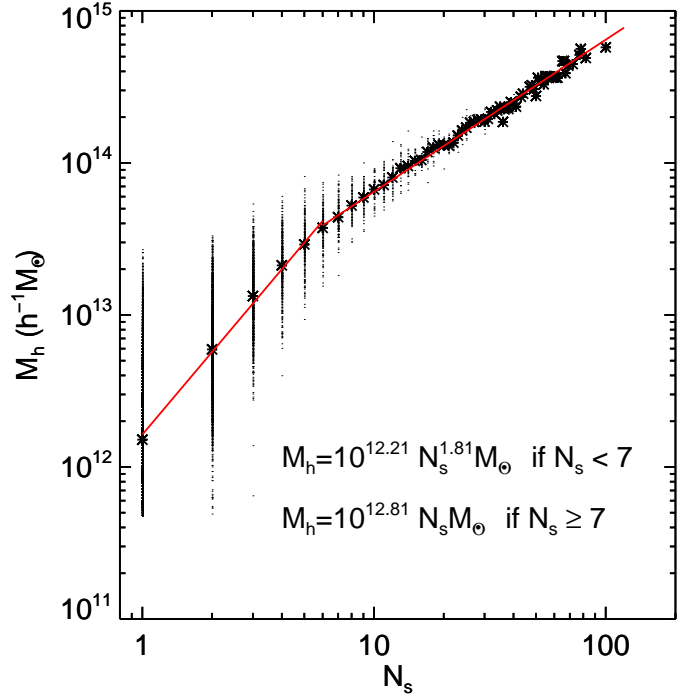}
\caption{The correlation between the  stellar-mass-defined  group richness$\Ns$
and host halo mass. The dashed line shows the fitting relation of equation
\ref{Equ_fitNsMh}.} \label{Fig_NsMh}
\end{figure}

As given redshift $z$, for the SDSS main galaxy sample, the galaxies are
volume complete at the stellar mass limit $\Log[M_{\rm {\rm s,lim}}]$.
\citet{Bosch08} obtained a fitting relation of the stellar mass limit $M_{\rm
  s,lim}$ as function of redshift $z$,
\begin{equation}\label{Equ_Msz}
\Log[M_{\rm {\rm s,lim}}/(h^{-2}M_\odot)]=\frac{4.852+2.246\Log
D_L(z)+1.123\Log(1+z)-1.186z}{1-0.067z}
\end{equation}
where $D_L(z)$ is the luminosity distance at redshift $z$. For us, we select
the groups with redshift $z<0.075$ and so that obtain a volume complete
sample of members down to stellar mass $M_{\rm s} \sim 3\times 10^{10} h^{-2}
M_\odot$. The stellar-mass-defined group richness $N_{s}$ is accordingly
defined as the number of members with stellar mass higher than $3\times
10^{10} h^{-2} M_\odot$. We show the correlation between $N_{s}$ and host
halo mass $\Mh$ in Fig. \ref{Fig_NsMh}.

We also fit a broken power-law relation between $\Mh$ and $\Ns$ with the break
point at $\Ns=7$. For high richness groups($\Ns \geq 7$), we fix the power law
index $n=1$. While for low richness groups($\Ns < 7$), we get the power law
index $n=1.81$ from a least square fitting of the linear relation between
$\Log\Mh$ and $\Log\Ns$. The resulted fitting formula is shown as equation
(\ref{Equ_fitNsMh}) and plotted as the solid line in Fig.  \ref{Fig_NsMh}.

\begin{equation}\label{Equ_fitNsMh}
  \Mh=\left\{
\begin{array}{ll}
   10^{12.21}\, \Ns^{1.81}\, \Msun &     \mbox{if $\Ns < 7$}\\
   10^{12.81}\, \Ns\,  \Msun       &     \mbox{if $\Ns \geq 7$}
\end{array}
\right.
\end{equation}

\begin{figure}
\centering
\includegraphics[width=85mm]{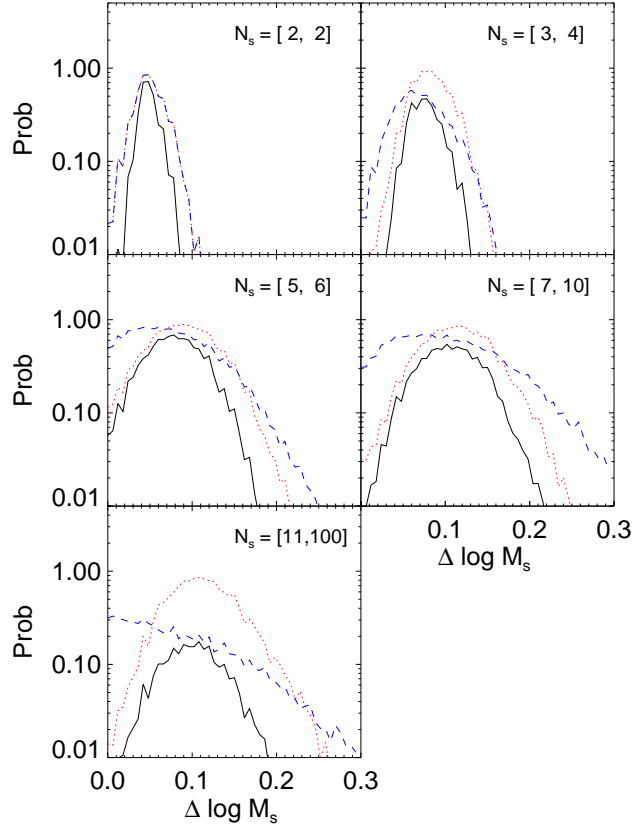}
\caption{The KS test probabilities $P_{\rm MGG}$, $P_{\rm SMG}$ and $P_{\rm M,S}$ as function
 of the average stellar mass increment $\Delta \Log M_s$   for 6 different $\Nrich$ bins, which are shown as
 the dotted, dashed and solid lines respectively in each panel.}
\label{Fig_MMG}
\end{figure}

As for BGGs, we test the model that the most massive galaxies(MMGs) of groups are initially
a statistical population, and then evolved with their stellar massed increased. We divide the
group sample into 5 $N_s$ bins.  The ranges of the 5 $\Ns$ bins are [2,2], [3,4], [5,6], [7,10],
[11,100] respectively. We also parameterize the average stellar mass increment of MMGs in
each richness bin with a simple parameter $\Delta \Log M_s$. As that in Section
\ref{Sec_Cmodel}, we also require $\Delta M_s$ can not be larger than the stellar mass
difference between the MMG and the second most massive galaxy.  We use KS test to check
whether the distribution of the stellar masses of the MMGs and the second most massive
galaxies are consistent with the  `shuffled' groups as that done in Fig. \ref{Fig_modelBS}. We
show the KS test probabilities  $P_{\rm MMG}$, $P_{\rm SMG}$ and the combined
probability ($P_{\rm M,S}\equiv P_{\rm MMG}*P_{\rm SMG}$)  as functions of  $\Delta \Log
M_s$ for 5 $\Ns$ bins in Fig. \ref{Fig_MMG}.

As for BGGs, we take the point where the accumulated probability from $\Delta \Log
\bar{M}_s=0$ is half of the accumulated probability  that from $\Delta \Log \bar{M}_s=0$ to
0.3 as the best estimation of the $\Delta \Log M_s$.  The best estimations of the average
stellar mass increment($\Delta \Log \bar{M}_s$) in logarithm space are [0.05, 0.08, 0.08,
0.10,0.11] dex for five richness bins respectively. The uncertainties of  $\Delta \Log
\bar{M}_s$ are also then calculated from the 68 percentiles of the likelihood distribution on
each side. The  $\Delta \Log \bar{M}_s$  together with their uncertainties  are plotted in the
top panel of Fig. \ref{Fig_DeltaM} and are listed in Table 1.

\end{document}